\documentclass[aps,prl,showpacs,floatfix,amsmath,amssymb,amsfonts,twocolumn,superscriptaddress]{revtex4-2}
\usepackage{graphicx}
\usepackage{color}
\usepackage{ulem}
\usepackage{bm}
\usepackage[colorlinks=true,allcolors=blue]{hyperref}
\usepackage{latexsym}
\usepackage{dsfont}
\usepackage{epstopdf}
\usepackage{CJK}
\usepackage{float}
\usepackage{titlesec}
\usepackage{physics}
\usepackage{natbib}

\begin{document}
\title{Collective excitations in the tetravalent lanthanide honeycomb antiferromagnet, Na$_2$PrO$_3$}

\author{M.~J.~Daum}
\thanks{These two authors contributed equally}
\affiliation{School of Physics, Georgia Institute of Technology, Atlanta, GA 30332, USA}

\author{A.~Ramanathan}
\thanks{These two authors contributed equally}
\affiliation{School of Chemistry and Biochemistry, Georgia Institute of Technology, Atlanta, GA 30332, USA}

\author{A. I. Kolesnikov}
\affiliation{Neutron Scattering Division, Oak Ridge National Laboratory, Oak Ridge, TN 37831, USA}

\author{S. A. Calder}
\affiliation{Neutron Scattering Division, Oak Ridge National Laboratory, Oak Ridge, TN 37831, USA}

\author{M.~Mourigal}
\email{mourigal@gatech.edu}
\affiliation{School of Physics, Georgia Institute of Technology, Atlanta, GA 30332, USA}

\author{H.~S.~La Pierre}
\email{hsl@gatech.edu}
\affiliation{School of Chemistry and Biochemistry, Georgia Institute of Technology, Atlanta, GA 30332, USA}
\affiliation{Nuclear and Radiological Engineering and Medical Physics Program, School of Mechanical Engineering, Georgia Institute of Technology, Atlanta, GA 30332, USA}

\date{October 13, 2020}
\begin{abstract} 
    Thermomagnetic and inelastic neutron scattering measurements on Na$_2$PrO$_3$ are reported. This material is an antiferromagnetic honeycomb magnet based on the tetravalent lanthanide Pr$^{4+}$ and has been proposed to host dominant antiferromagnetic Kitaev interactions. These measurements reveal magnetic fluctuations in Na$_2$PrO$_3$ below an energy of 2~meV as well as crystal-field excitations around 230~meV. The latter energy is comparable to the scale of the spin-orbit interaction and explains both the very small effective moment of around $1.0~\mu_{\rm B}$ per Pr$^{4+}$ and the difficulty to uncover any static magnetic scattering below the ordering transition at $T_{\rm N}= $ 4.6~K. By comparing the low-energy magnetic excitations in Na$_2$PrO$_3$ to that of the isostructural spin-only compound, Na$_2$TbO$_3$, a microscopic model of exchange interactions is developed that implicates dominant and surprisingly large Heisenberg exchange interactions $J\approx1.1(1)$~meV. Although antiferromagnetic Kitaev interactions with ${\rm K}\leq0.2J$ cannot be excluded, the inelastic neutron scattering data of Na$_2$PrO$_3$ is best explained with a $\Delta = 1.24(2)$ easy-axis XXZ exchange anisotropy. 
\end{abstract}
\maketitle

Frustrated quantum magnets have been proposed as a platform to realize quantum spin-liquids (QSLs) and other exotic forms of magnetic matter~\cite{Balents2010, Broholm2020}. In QSLs, quantum fluctuations are so strong that spins remain disordered for temperatures well below the average interaction scale between spins and become entangled. Geometrically frustrated lattices featuring  lanthanide ions have gained much recent attention, including the triangular, kagome, and pyrochlore systems~\cite{sibille2015candidate,Kimura2013,Paddison2017,bordelon2019field,dun2020quantum,sibille2020quantum}. An alternative realization of a QSL (with an exact solution) was proposed by Kitaev based on $S\!=\!1/2$ moments on a honeycomb lattice with bond-dependent Ising-like interactions~\cite{Kitaev2006}. While the honeycomb lattice is not inherently frustrated, anisotropic interactions, parametrized by the Kitaev term (${\rm K}$), give rise to frustration between the competing orthogonal anisotropy axes. However, in real materials, the Kitaev interactions are often perturbed by Heisenberg interactions ($J$) giving rise to the strongly frustrated Heisenberg-Kitaev ($J$--${\rm K}$) model~\cite{Chaloupka2010, Singh2012}.

\begin{figure}[t!]
	\includegraphics[width=1.0\columnwidth]{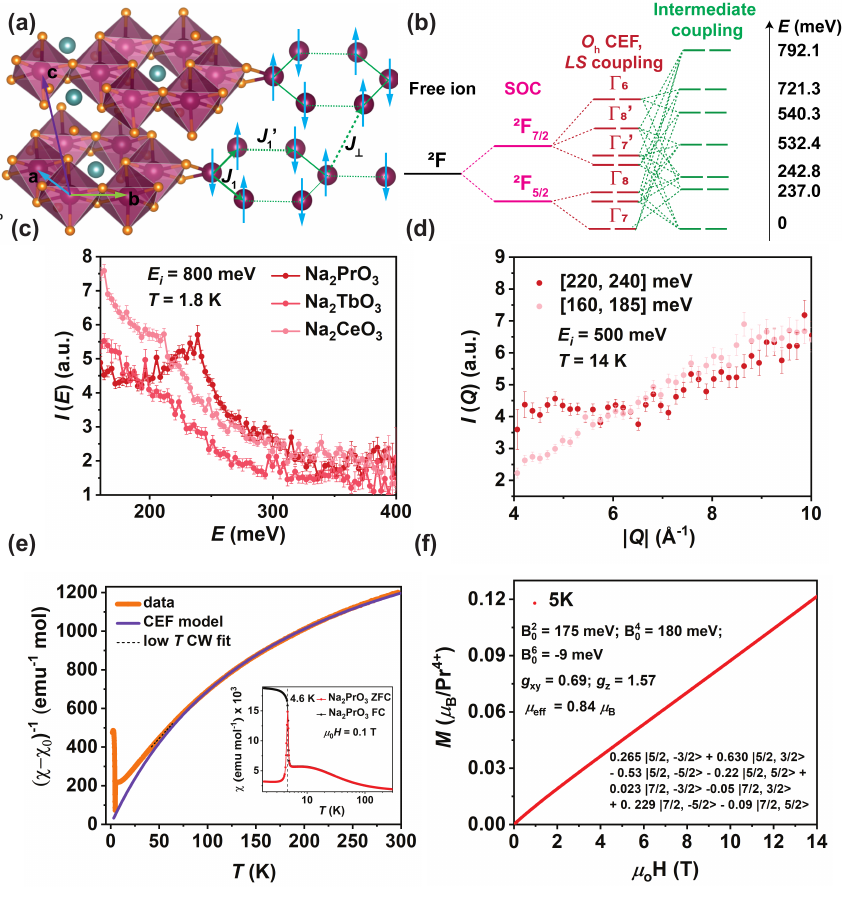}
	\caption{(a) Monoclinic crystal structure of Na$_2$PrO$_3$ showing the honeycomb layers of Pr$^{4+}$ ions and the resulting nearest neighbor and out-of-plane magnetic exchange pathways. (b) Comparison of the expected splitting of the $^2F_{5/2}$ free-ion ground-state of Pr$^{4+}$ by SOC and a $O_h$ CEF environment in $LS$ and intermediate coupling. (c) Comparison between low momentum-transfer $Q\leq6$~\AA$^{-1}$ inelastic neutron scattering spectra measured with $E_i\!=\!800$ meV for Na$_2$$Ln$O$_3$ (${Ln}$ = Ce, Pr, Tb) at $T\!=\!1.8$ K. (d) Momentum-transfer dependence of the {$E\!=\!233(1)$~meV} excitation at $T=14$~K compared to phonon background at comparable energies. (e) The inverse magnetic susceptibility $1/(\chi(T)-\chi_0)$ and susceptibility $\chi(T)$ of Na$_2$PrO$_3$ in a $0.1$~T field and $\chi_0=1.09\times10^{-3}$ emu mole$^{-1}$. The bold solid line (purple) is the CEF comparison to the data with the parameters of panel (f). The black and red traces in the inset corresponds to field-cooled (FC) and zero-field cooled measurements, respectively. (f) Isothermal magnetization $M(H)$ at $T\!=\!5$ K and obtained CEF parameters. }  
	\label{fig1}
\end{figure}

Bond-dependent interactions stem from strong spin-orbit coupling in magnetic insulators. Hence heavy $4d$ and $5d$ transition metal ions have been proposed as a paradigm to realize a Kitaev QSL~\cite{Motome2020, Takagi2019}. Spin-orbit $J_{\rm eff} = 1/2$ Mott insulators comprising low-spin $d^5$ and $d^7$ transition metal ions, such as Na$_2$IrO$_3$, H$_3$LiIr$_2$O$_6$, Li$_2$IrO$_3$, and RuCl$_3$ have been extensively studied to search for Kitaev physics \cite{Banerjee2016, Banerjee2017, Choi2012, Kitagawa2018, Takayama2015, Sano2018, Liu2018}. An alternative approach is to explore $J_{\rm eff}\!=\!1/2$ magnetic moments from $f$-element ions, which exhibit significant anisotropy~\cite{Li2017, Rau2018}. In the $4f$ electron manifold, several electron configurations can host $J_{\rm eff}\!=\!1/2$ magnetic moments, with the one electron/one-hole ($4f^1/4f^{13}$) configurations as the most desirable. The one hole case is realized by the $J_{\rm eff}=1/2$ honeycomb material YbCl$_3$, the collective behavior of which was recently shown to be best described from the Heisenberg limit \cite{Sala2020}. The one electron case leads to the $4f^1$ ions Ce$^{3+}$ or Pr$^{4+}$ including Na$_2$PrO$_3$, a material with edge sharing PrO$_6$ octahedra forming a honeycomb network similar to the iridates and recently proposed to exhibit dominant antiferromagnetic Kitaev interactions, contrasting with $4d$ and $5d$ systems~\cite{Jang2019}.

In this work, the magnetic properties of Na$_2$PrO$_3$ are investigated using a combination of thermomagnetic and neutron scattering measurements on powder samples. These studies uncover spin-wave-like excitations at energies below 2~meV. A comparison to the isostructural compounds Na$_2$TbO$_3$ and Na$_2$CeO$_3$, that represent spin-only magnetic-moment and non-magnetic analogues of the title compound, respectively, yield deeper insights into the effective magnetic Hamiltonian of Na$_2$PrO$_3$. Although no magnetic Bragg peaks are observed within experimental sensitivity below the $T_{\rm N}\approx4.6$~K transition seen in thermomagnetic probes, dynamic correlations in Na$_2$PrO$_3$ are well explained by a model including antiferromagnetic nearest-neighbor interactions and a easy-axis XXZ exchange anisotropy. The inelastic data does not support the presence of a sizeable Kitaev term ${\rm K}$. These studies also reveal an unusually small effective magnetic moment for the Pr$^{4+}$ ions which is explained by the increased crystal-field splitting in comparison to trivalent lanthanides \cite{gompa2020chemical, minasian2017quantitative}.

Polycrystalline samples of Na$_2${\it Ln}$^{4+}$O$_3$ with ${Ln}$ = Ce (4$f^0$), Pr (4$f^1$), Tb (4$f^7$) were synthesized by solid-state reactions and structurally characterized by synchrotron x-ray diffraction [see Supplementary Information (SI) Sec.~S1]. Na$_2$PrO$_3$ contains layers of PrO$_6$ octahedra forming distorted honeycomb networks separated by layers of Na ions, with two intraplane Pr--Pr distances $d=3.407(3)$~\AA\, and $d^\prime=3.487(6)$~\AA, and an inter-plane distance of $d_\perp\!\approx\!5.8$~\AA\, at $T=100$~K [Fig.~\ref{fig1}(a)]. The ABC stacking sequence in the $C2/c$ space group originates from symmetry breaking displacements of the Na atoms which also lead to evident stacking faults in diffraction patterns. Na$_2$CeO$_3$ and Na$_2$TbO$_3$ are isostructural to Na$_2$PrO$_3$~\cite{RN204} [see SI Sec.~S1]. Given the air-sensitivity of these samples, all synthesis and measurement operations where performed in an inert-gas atmosphere.

To understand the single-ion properties of Na$_2$PrO$_3$, broadband inelastic neutron scattering measurements on the fine-resolution Fermi chopper spectrometer (SEQUOIA)~\cite{Granroth2010,Stone2014} were performed at the Spallation Neutron Source (SNS), Oak Ridge National Laboratory (ORNL). Experiments were performed on ($m\!=\!8$~g) polycrystalline samples loaded in annular Al powder cans and inserted into a liquid $^4$He cryostat reaching a base temperature of $T\!=\!1.5$~K. The data was reduced in {\scshape Mantid}~\cite{Arnold2014} to yield the neutron scattering intensity $I(Q,E)$ as a function of momentum-transfer $Q$ and energy-transfer $E$. We used a series of incoming energies to probe possible crystal electric field (CEF) excitations of our samples up to an energy transfer of $E\!\approx\!500$~meV [see SI Sec.~S2]. 

Pr$^{4+}$ is a $4f^1$ Kramers ion, isoelectronic to Ce$^{3+}$, with a $^2F_{5/2}$ free-ion ground-state. For an octahedral oxygen environment with $O_h$ symmetry, the CEF splitting leads to a Kramers doublet ground-state ($\Gamma_7$) and an excited quartet $(\Gamma_8)$ which we expect to split into two doublets given the lower $D_{2d}$ site-symmetry of Pr$^{4+}$ in Na$_2$PrO$_3$ [Fig.~\ref{fig1}(b)]. The energy-dependence of the neutron scattering intensity at low momentum-transfer $I(Q\!\leq\!6$~\AA$^{-1},E)$ was used to search for these CEF excitations. The comparison of different $E_i$'s and samples [Fig.~\ref{fig1}(c) and SI Sec.~S2] reveals a strong excitation at $E\!=\!233(1)$~meV. The intensity of the excitation  increases at low $Q$ as expected for magnetic scattering [Fig.~\ref{fig1}(d)]. The excitation found in Na$_2$PrO$_3$ compares well to the 260~meV $\Gamma_7$ to $\Gamma_8$ splitting observed for BaPrO$_3$~\cite{Kern1985} which comprises Pr$^{4+}$ ions in an ideal $O_h$ environment. Since no other CEF excitations are observed below $500$~meV, the $E\!= \!233$~meV mode is associated with the two quasi-degenerate $\Gamma_8$ doublets [See SI Sec.~S2]. 

Given this quasi-degeneracy, the CEF Hamiltonian can be written using the Wybourne tensor operators as $\mathcal{\hat{H}}_{\rm CEF} = B^2_0 \hat{C}^2_0 + B^4_0 (\hat{C}^4_0+5\hat{C}^4_4 ) + B^6_0 (\hat{C}^6_0- 21\hat{C}^6_4)$ where the $B^2_0$  parameter reflects the small axial distortion of the PrO$_6$ octahedral away from $O_h$ [See SI Sec.~S3]. The large CEF energy scale in Na$_2$PrO$_3$ has been observed indirectly by O $K$-edge X-ray absorption near edge spectroscopy studies of PrO$_2$~\cite{minasian2017quantitative}, and is similar in magnitude to the  spin-orbit interaction $\lambda\!\approx\!100$~meV resulting in a $\approx360$~meV separation between $^2F_{5/2}$ and $^2F_{7/2}$ for a free Pr$^{4+}$ ion \cite{kaufman1967fifth, hinatsu1994electron}. As a result, a mixing of the $^2F_{5/2}$ and $^2F_{7/2}$ electronic manifolds is expected and the above Hamiltonian must be diagonalized using the complete set of intermediate-coupling basis states using {\scshape Spectre} \cite{boothroydspectre}. 

The three CEF parameters are constrained by the observed excitation and further determined by matching the calculated temperature-dependence of the magnetic susceptibility to our measurements for $\mu_0H\!=\!0.1$~T [Fig.~\ref{fig1}(e)]. An excellent agreement is obtained for $T\geq60$~K for the CEF parameters of Fig.~\ref{fig1}(f) and a temperature independent term $\chi_0$ is determined to be $1.09 \times 10^{-3}$ emu.mol$^{-1}$. The value of $\chi_0$ is in reasonable agreement with the estimated value~\cite{Chaloupka2013} $\chi_0 \!\approx\! \frac{8}{3\lambda} \mu^{2}_{\rm B} N_{\rm A} = 0.8 \times 10^{-3}$ emu.mol$^{-1}$ for a Pr$^{4+}$ ion and with observations for the related compound BaPrO$_3$ \cite{bickel1988magnetic}. This result indicates a ground-state doublet  dominated by $|^2F_{5/2},\pm3/2\rangle$ and $|^2F_{7/2},\pm5/2\rangle$ states and predicts several higher-energy doublets beyond the $500$~meV reach of our experiments [see SI Sec.~S3]. This fit also yields a calculated powder-averaged $g\!=\!0.98$ for effective $J_{\rm eff}\!=\!1/2$. The effective moment $\mu^{\rm eff}_{\rm CEF}\!=\!0.84~\mu_{\rm B}/$Pr is unusually small given the free-ion value is $\mu_{\rm eff}^{\rm free} = 2.54 \mu_{\rm B}/$Pr. The small $g$-tensor is evident from the experimental isothermal magnetization at $T\!=\!5$~K, which is linear and reaches only $\approx 0.12~\mu_{\rm B}/$Pr at $\mu_0{H}\!=\!14$~T, far short of saturation [Fig.~\ref{fig1}(f)].

\begin{figure}[t!]
	\includegraphics[width=1.0\columnwidth]{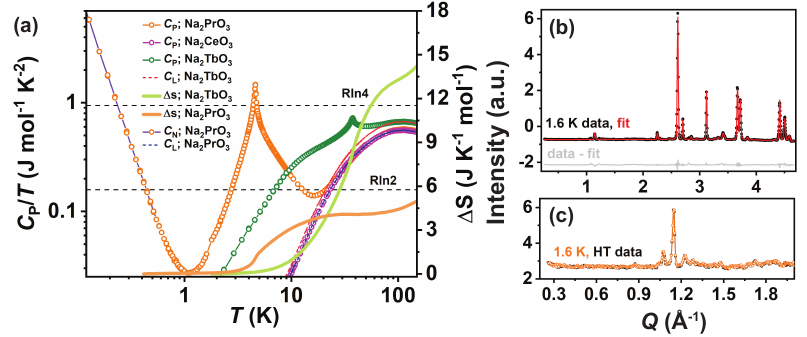}
	\caption{(a) Heat capacity measurments of our Na$_2${\it Ln}$^{4+}$O$_3$ samples measured using the relaxation method above $T=2$~K. For Na$_2$PrO$_3$, measurements down to $T\!=\!100$~mK were carried out using a dilution refrigerator insert on pressed pellets mixed with Ag. The change in magnetic entropy for Na$_2$PrO$_3$ and Na$_2$TbO$_3$ is obtained after subtracting the (scaled) lattice contribution from Na$_2$CeO$_3$. (b) Neutron diffration measurements on Na$_2$PrO$_3$ at $T=1.6$~K using $\lambda=2.4$~\AA. The red line is the result from a Rietveld refinement. (Inset) Comparison at low-angle diffraction between $T=1.6$~K and $T=20$~K.} 
	\label{fig2}
\end{figure}

Below $T\approx40$~K, the susceptibility of Na$_2$PrO$_3$ deviates from the single-ion form and culminates in a magnetic transition at $T_N\!=\!4.6$~K, consistent with Ref.~\onlinecite{Hinatsu2006}, with a clear splitting between field-cooled (FC) and zero-field cooled (ZFC) traces but no visible frequency dependence in ac susceptibility [See SI Sec.~S4]. Thus, this sharp peak is interpreted as magnetic ordering preceded by short-range order [Fig.~\ref{fig1}(e)]. It is difficult to find an adequate regime for a Curie-Weiss analysis: a fit limited to $40\leq T \leq60$ K yields an antiferromagnetic Weiss constant {$\Theta_{\rm W}\!=\!-30.4(1)$~K} and {$\mu^{\rm eff}_{\rm CW} = 1.19(1)\mu_{\rm B}/$Pr}, comparable to $\mu^{\rm eff}_{\rm CEF}$. Heat capacity measurements [See SI Sec.~S5] corroborate this picture [Fig.~\ref{fig2}(a)]. An additional upturn is observed below $T=0.3$~K, which is associated with nuclear spins. Subtracting the lattice contribution reveals an entropy change {$\Delta S \approx0.76 R \ln 2$} between 0.1~K and 40~K, corroborating the $J_{\rm eff}\!=\!1/2$ picture for Pr$^{4+}$ and revealing some missing entropy.

\begin{figure*}[t!]
                \includegraphics[width=0.88\linewidth]{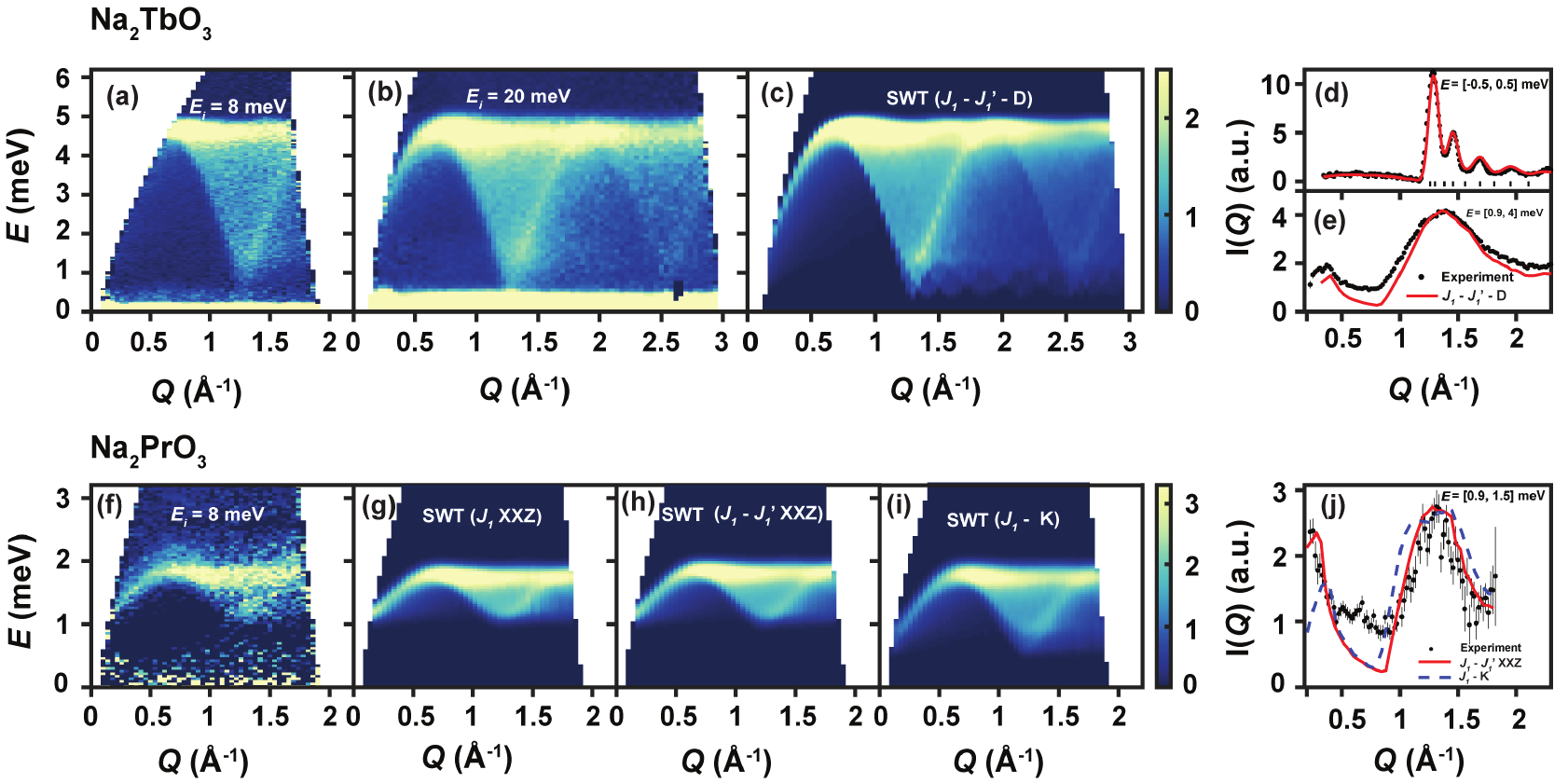}
                \caption{(a-e) Inelastic neutron scattering intensity $I(Q,E)$ from Na$_2$TbO$_3$ at $T=1.5$~K using (a) $E_i=8$~meV or (b)  $E_i=20$~meV. (c) Comparison with powder-averaged linear spin-wave theory calculations for optimized parameters including the magnetic form factor of Tb$^{3+}$ and a constant energy energy broadening factor of 0.2~meV. (d) Elastic magnetic scattering $-0.5\leq E \leq0.5$~meV of  Na$_2$TbO$_3$ obtained by subtracting $T=20$~K data from {$T=55$~K} (black dots). The red line and vertical black ticks are the result of a Rietveld refinement using a $\boldsymbol{k}_{\rm m}\!=\!0$ propagation vector. (e) Momentum dependence of the low-energy inelastic signal $0.9\leq E \leq4$~meV (black dots) and comparison to linear spin-wave-theory predictions for optimized parameters (red line). (f-j) Inelastic neutron scattering results from Na$_2$PrO$_3$ at $T=1.5$~K using (f) $E_i=8$~meV and $T=55$~K subtracted. Comparison to linear spin-wave-theory predictions with optimized parameters for (g) a $J_1$ XXZ model with $\Delta=1.22$ ($J_1\!=\!1.06$~meV), (h) a $J_1$--$J_1^\prime$ XXZ model with $\Delta=1.26$ and $ J_1^\prime=0.85 J_1$ ($J_1\!=\!1.1$~meV), and (i) a $J_1$--${\rm K}$ model with $K=0.18 J_1$ ($J_1 = 1.1$~meV). (i) Momentum dependence of the low-energy inelastic signal $0.9\leq E \leq1.4$~meV (black dots) and comparison to linear spin-wave-theory predictions for optimized parameters from the $J_1$--$J_1^\prime$ XXZ (solid red line) and $J_1$--${\rm K}$ models (dashed blue line). To avoid over subtraction, the cut is taken from empty cryostat subtracted data while (f) shows a temperature subtracted spectrum.}
                \label{fig3}
\end{figure*}

To understand the ground-state of Na$_2$PrO$_3$ below the transition, neutron powder diffraction experiments were performed on the HB2A diffractometer~\cite{Garlea2010} at the High Flux Isotope Reactor (HFIR), ORNL. No additional Bragg peaks are observed beyond the $C2/c$ nuclear structure [Fig.~\ref{fig2}(b)], even after subtracting a {$T=55$}~K background [Fig.~\ref{fig2}(b)-inset]. Given the high incoherent scattering background from the sample [See SI Sec.~S2], the small effective moment of Pr$^{4+}$, the stacking faults in the crystal structure, and the likelihood of a $\boldsymbol{k}_{\rm m}\!=\!0$ propagation vector, this result is not entirely surprising. To get an estimate on any ordered moment $\langle\mu^z\rangle$, the hyperfine coupling in the nuclear specific-heat was modeled using a Schottky form. Assuming the entire upturn is nuclear yields a static electronic moment $\langle \mu^z_{\rm hyp}\rangle = 0.41\mu_{\rm B}$ at the time-scale of the nuclear-lattice relaxation~\cite{kimura2013quantum}, which is comparable to $\langle \mu^z_{\rm CEF} \rangle = 0.49~\mu_{\rm B}$ estimated from CEF calculations.

In the absence of visible magnetic Bragg peaks in Na$_2$PrO$_3$, low-energy inelastic neutron scattering ($E_i\!=\!8$, $20$ meV) was employed to search for magnetic fluctuations. It is instructive to compare these results to the isostructural spin-only compound, Na$_2$TbO$_3$ [Fig.~\ref{fig3}(a--e)]. Na$_2$TbO$_3$ orders at ${T}^\prime_{\rm N}=38.2$~K and develops structured spin-wave excitations below ${T}^\prime_{\rm N}$ [Fig.~\ref{fig3}(a,b)] with a band-top of $5$~meV and a $\approx1$~meV gap. Inspecting the elastic line, and subtracting $T=55$~K data from $T=10$~K, evidences intense magnetic Bragg peaks indexed by a $\boldsymbol{k}_{\rm m}\!=\!0$ propagation vector [Fig.~\ref{fig3}(d)]. Representation analysis in {\scshape sarah}~\cite{Wills_2000} yields four possible magnetic structures, only one of which yields a good fit following a Rietveld refinement in {\scshape fullprof}~\cite{Rodriguez-Carvajal_1993} [See SI Sec.~S6]. Spins in the resulting N\'eel ordered structure [Fig.~\ref{fig3}(d)-inset] lie in the $ac$-plane, essentially along $c$. Spin-wave excitations in Na$_2$TbO$_3$ are very intense given $S=7/2$ and can be efficiently modeled using linear spin-wave theory~\cite{Petit2011} in {\scshape spinW}~\cite{Toth2014}. A spin Hamiltonian $\mathcal{H} = \mathcal{H}_{\rm ex} +D \sum_i (S_i^z)^2$ with Heisenberg exchange interactions $J_1$ and $J_1^\prime$ for the split nearest-neighbor pairs ($d$ and $d^\prime$), $J_\perp$ between inequivalent lanthanide sites in two adjacent honeycomb planes, and a single-ion anisotropy term $D$ is considered. The calculated powder-averaged intensity is in excellent agreement with the data [Fig.~\ref{fig3}(c)] with parameters obtained after a grid-calculation and subsequent search for a minimal $\chi^2$ [See SI Sec.~S7]. These parameters, $J_1\!=\!0.50$ meV, $J_1^\prime\!=\!0.85 J_1$, $J_{\perp}\!=\!-0.02J_1$ and $D\!=\!-0.001J_1$, indicate that the observed band-top dispersion [Fig.~\ref{fig3}(a,b)] is induced by the splitting of $J_1$ and $J_1^\prime$, and that the small spin gap is the combined effect of ferromagnetic $J_{\perp}$ and easy-axis $D$.

This forgoing analysis facilitates the description of the magnetic fluctuations in Na$_2$PrO$_3$ at $T\!=\!1.5$~K [Fig.~\ref{fig3}(f--i)], which resemble the spin-wave excitations in Na$_2$TbO$_3$ [Fig.~\ref{fig3}(a)], but with a reduced band-top of $2$~meV [Fig.~\ref{fig3}(f)] and a ten-fold decrease in scattering intensity. Thus, a high temperature subtraction ($T=55~K$) was utilized [Fig.~\ref{fig3}(f)]. Given that no static magnetic scattering is observed, an incipient $\boldsymbol{k}_{\rm m}\!=\!0$ order is assumed by analogy with Na$_2$TbO$_3$. Given the $J_{\rm eff}=1/2$ magnetic moments, several models are adopted to incorporate exchange anisotropies (scaled to the corresponding primary exchange) on $J_1$ and $J_1^\prime$ bonds, with spin-wave theory calculations performed in {\scshape spinW} [See SI Sec.~S8]. Including a diagonal exchange anisotropy (XXZ), \textit{e.g.} $\mathcal{H}_{\rm ex} \equiv J \sum_{ij} ({S}^x_i {S}^x_j + {S}^y_i {S}^y_j + \Delta {S}^z_i {S}^z_j)$,  opens a gap in the spectrum and yields an excellent agreement with the data for $J_1 \!=\!J_1^\prime\!=\! 1.06$ meV and $\Delta=1.22$ [Fig.~\ref{fig3}(g)]. Allowing $J_1^\prime$ to vary independently slightly broadens the bandwidth but does not significantly improve the agreement between data and calculations [Fig.~\ref{fig3}(h)]. Introducing a Kitaev term ${\rm K}$ in the Hamiltonian yields an overall agreement with the data for $J_1 = 1.1$~meV ($J_1^\prime\!=\!J_1$) and an antiferromagnetic ${\rm K}\!=\!0.18 J_1$, but introduces a weak double-gap feature at the band bottom that is clearly not observed in the experiment [Fig.~\ref{fig3}(i)]. A cut through the low-energy part of the data and the corresponding spin-wave calculations [Fig.~\ref{fig3}(j)] shows overall agreement, except between $Q=0.25$~\AA$^{-1}$ and $1$~\AA$^{-1}$ where background contributions are large. All models fail to account for the apparent continuum at the top of the band, which we attribute to the presence of quantum fluctuations. Although single-crystal studies will be necessary to determine the relative importance of these terms in detail.

In conclusion, the $1.1(1)$~meV energy scale of the Heisenberg exchange interaction in Na$_2$PrO$_3$ is surprisingly large for a lanthanide system, that is reflected in the large $230$~meV scale of the crystal-electric field and the necessity to employ an intermediate-coupling scheme to explain the small effective moment of $1.0 \mu_{B}$ per Pr$^{4+}$. In turn, these effects are hypothesized to be the origin of the considerably weaker antiferromagnetic Kitaev interaction, extracted by modeling the magnetic fluctuations of Na$_2$PrO$_3$, when compared to theoretical and \textit{ab-initio} calculations~\cite{Jang2019}. The absence of visible magnetic Bragg peaks in Na$_2$PrO$_3$ as well as the missing entropy of around $0.3 R\ln2$ are two avenues for future inquiry. In particular, the observed low moment for the Pr$^{4+}$ ion has important implications for the understanding and application of high-valence lanthanide ions in magnetic materials, since, akin to observations in high-valence actinides, this moment is derived from competition between SOC and CEF and necessitates the use of an intermediate coupling scheme to capture the observed temperature dependence~\cite{ magnani2005perturbative, escalera2019exploring}.

\acknowledgements

We thank Zhiling Dun for numerous helpful discussions. The work of A.R. and H.S.L. at Georgia Tech was supported by the Beckman Foundation as part of a Beckman Young Investigator Award to H.S.L. The work of M.D. and M.M. at Georgia Tech was supported by the National Science Foundation through grant NSF-DMR-1750186. This research used resources at the High Flux Isotope Reactor and Spallation Neutron Source, a DOE Office of Science User Facility operated by the Oak Ridge National Laboratory.

%

\end{document}